\documentclass[12pt]{article}

\begin{document}

\begin{titlepage}

\topmargin=3.5cm

\textwidth=13.5cm

\centerline{\Large \bf Nonabelian Localization for   }

\vspace{0.4cm}

\centerline{\Large \bf  Statistical Mechanics of Matrix Models}

\vspace{0.4cm}
\centerline{\Large \bf at High Temperatures}

\vspace{1.0cm}

\centerline{\large \bf Levent Akant\footnote{E-mail:
akant@gursey.gov.tr}}

\vspace{0.5cm}

\centerline{ \textit{Feza Gursey Institute}}

\centerline{\textit{Emek Mahallesi, Rasathane Yolu No.68 }}
\centerline{\textit{Cengelkoy, Istanbul, Turkey}}

\vspace{1.5cm}

We show that in the high temperature limit the partition function of a matrix model is localized on certain
shells in the phase space where on each shell the classically conjugate matrix variables obey the canonical commutation relations. The result is obtained by applying the nonabelian equivariant localization principle to
the partition function of a matrix model driven by a specific random external source coupled to a conserved charge of the system.  

\end{titlepage}

\pagebreak

\baselineskip=0.6cm

\abovedisplayskip=0.4cm

\belowdisplayskip=0.4cm

\abovedisplayshortskip=0.3cm

\belowdisplayshortskip=0.3cm

\jot=0.35cm

\textheight=20cm

\section{Introduction}

Recently there have been attempts to derive quantum dynamics (canonical commutation relations and Heisenberg equations of motion) as an emergent phenomenon. Interesting results in this problem are obtained by S. Adler and collaborators in a series of papers \cite{Adler1, Adler2, Adler3, Adler-Horwitz} where the quantum dynamics is derived, under certain approximations and assumptions, from the classical statistical mechanics of matrix models. 

Here we want to present a result which might be useful in clarifying certain aspects of this emergent phenomenon. 
We consider the statistical mechanics of a $0+1$ dimensional Hermitean matrix model in the presence of a random external source with a Gaussian distribution and show that in the high temperature limit the conjugate variables (conjugate with respect to the Poisson brackets) obey the canonical commutation relations. 
Since this result is obtained in the high temperature limit it is independent of the specific form of the Hamiltonian of the system. Thus the only assumption we make about the structure of the matrix models is the randomness of the external source.
We derive our result by using Witten's nonabelian localization principle \footnote{The adjective nonabelain refers to the noncommutativity of a symmetry group. The abelian localization was derived by Duistermaat and Heckman \cite{DH} } \cite{Witten1, Witten2, Witten3, KJ}. We show that the term in the partition function which survives the high temperature limit is a closed equivariant form. Then we use the nonabelian localization principle to show that the partition function is localized on a certain subset of the phase space. The subset in question is characterized by a 2 dimensional Yang-Mills algebra. This algebra is an example of a cubic Artin-Schelter algebra and is the universal enveloping algebra of the Heisenberg algebra. 

In Sec. 2 we construct the phase spaces of matrix models that we will consider. We show that 
the unitary group $U(N)$ has a Hamiltonian action on the phase space and derive the corresponding moment map which is conserved whenever the Hamiltonian is $U(N)$ invariant.

In Sec. 3 we work out the statistical mechanics of matrix models defined in Sec. 2. Since the moment map is
a conserved quantity for $U(N)$ invariant systems we expect the Boltzmann factor to depend not only on the 
the Hamiltonian but also on the moment map. This dependence on the moment map is responsible for a nontrivial high temperature limit of the partition function. We assume that the external source coupling to the moment map
is itself random with a Gaussian distribution (white noise). This assumption allows us to apply the nonabelian
localization to the partition function in the high temperature limit.

In Sec. 4 we review the basics of equivariant cohomology and nonabelian localization principle.

Finally in Sec. 5 we apply the nonabelian localization principle to the partition function derived in Sec. 3
and show that the latter is localized on a subset of the phase space characterized by the canonical commutation relations.

\section{Matrix Models}
A matrix model is a dynamical system whose configuration space is
a subset of the space of $N\times N$ real or complex matrices. In
most applications the configuration space is either a vector space
(for example hermitean matrix models) or a group manifold (for
example unitary matrix models). In this article we will work with
hermitean matrix models. Thus the configuration space
$\mathcal{C}$ will be the space of all $N\times N$ hermitean
matrices. The phase space $\mathcal{M}=T^{*}\mathcal{C}$ of the
system will be identified with $\mathcal{C}^{2}$. The coordinates
in $\mathcal{M}$ will be denoted by $(X,P)$. We will also use the
notation $X_{1}=Re X$ and $X_{2}=Im X$, and similarly for $P$.

A symplectic form on $\mathcal{M}$ is defined as follows
\begin{equation}
\omega=\sum_{r=1}^{2}dX_{rij}\wedge dP_{rij}=Re\; Tr\; (dX\wedge
dP^{\dagger}).
\end{equation}
The action of the unitary group $U(N)$ given by
\begin{eqnarray}
X&\rightarrow & UXU^{-1} \\
P&\rightarrow & UPU^{-1}
\end{eqnarray}
is symplectic. 

The action of the fundamental vector field
corresponding to an element $-iA$ ($A^{\dagger}=A$) of
$\textit{u}(N)$ is given by
\begin{eqnarray}
\delta_{A}X&=&i\left[ A,X\right]\\
 \delta_{A}P&=&i\left[ A,P\right].
\end{eqnarray}
In terms of real and imaginary components this is equivalent to
\begin{eqnarray}
\delta_{A}X_{1}&=&-\left[ A,X\right]_{2}\\
 \delta_{A}X_{2}&=&\;\;\;\left[ A,X\right]_{1}.
\end{eqnarray}
and similarly for $P$. Thus the fundamental vector field itself is given by
\begin{equation}
\xi_{A}=\epsilon_{rs}\left( \left[
A,X\right]_{rij}\frac{\partial}{\partial
X_{sij}}+[A,P]_{rij}\frac{\partial}{\partial P_{sij}}\right ) .
\end{equation}
\newtheorem{sub}{Proposition}[section]

\begin{sub}
  The action of $U(N)$ on $\mathcal{M}$ is Hamiltonian.
  
\end{sub}
\textit{Proof:}
We must show that (i) the fundamental vector fields are Hamiltonian and (ii) that the corresponding moment maps
form a Lie algebra (with respect to the Poisson bracket induced by $\omega$) homomorphic to $\mathcal{G}$ 

(i) The fundamental vector fields are Hamiltonian i.e. there exist a function
$\mu_{A}$ on $\mathcal{M}$ such that
\begin{equation}
i_{\xi_{A}}\omega=\epsilon_{rs} \left[
A,X\right]_{rij}dP_{sij}-\epsilon_{rs} \left[
A,P\right]_{rij}dX_{sij}=-d\mu_{A}.
\end{equation}
Let
\begin{equation}
\mu_{A}=\; Tr\; (-iA\left[ X,P\right]).
\end{equation}
then
\begin{eqnarray}
  \;Tr\; (-iA\left[ X,P\right]) &=&  \; Im \;Tr\; (A\left[ X,P\right])  \\
   &=& - \; Im \;Tr\; (\left[ A,P\right]X) \\
   &=&  \; Im \;Tr\; (\left[ A,X\right]P)
\end{eqnarray}
but
\begin{eqnarray}
 \; Im \;Tr\; (\left[ A,P\right]X)&=& \;Tr\; ( \left[
A,P\right]_{1}X_{2}+\left[ A,P\right]_{2}X_{1})\\
&=& - \; \epsilon_{rs}\left[ A,P\right]_{rij}X_{sij}
\end{eqnarray}
thus
\begin{equation}
    -\frac{\partial \mu_{A}}{\partial X_{sij} }=-\epsilon_{rs}\left[
    A,P\right]_{rij}.
\end{equation}
Similarly,
\begin{equation}
-\frac{\partial \mu_{A}}{\partial P_{sij} }=\epsilon_{rs}\left[
    A,X\right]_{rij}.
\end{equation}
So we conclude
\begin{equation}
    -d\mu_{A}=i_{\xi_{A}}\omega.
\end{equation}

(ii) Let $-iT_{a}$ be a basis for $\textit{u}(N)$ with
$T_{a}^{\dagger}=T_{a}$, $[T_{a},T_{b}]=if_{abc}T_{c}$ and $Tr\;T_{a}T_{b}=\frac{1}{2}\delta_{ab}$. 

Define
\begin{eqnarray}
     \xi_{a}&\equiv& \xi_{T_{a}}= \epsilon_{rs}\left( \left[
T_{a},X\right]_{rij}\frac{\partial}{\partial
X_{sij}}+[T_{a},P]_{rij}\frac{\partial}{\partial P_{sij}}\right )\\
    \mu_{a}&\equiv& \mu_{\xi_{T_{a}}}=Tr \; (T_{a}[X,P]).
\end{eqnarray}
We want to show that the action of $\textit{u}(N)$ is equivariant
i.e.
\begin{equation}
    \{\mu_{a},\mu_{b}\}=if_{abc}\mu_{c}.
\end{equation}
Here $\{\;\;,\;\;\}$ is the Poisson bracket corresponding to the
symplectic form $\omega$. We have
\begin{eqnarray}
   \{\mu_{a},\mu_{b}\}&=& \omega(\xi_{a},\xi_{b})  \\
   &=& (i_{\xi_{a}}\omega)(\xi_{b}) \\
   &=&  Tr \; ([T_{a},P][X,T_{b}]+[T_{a},X][P,T_{b}])\\
   &=& Tr \; [T_{a},T_{b}][X,P]\\
   &=& if_{abc}\mu_{c}.
\end{eqnarray}

Hence we conclude that the action of $\textit{u}(N)$ on
$\mathcal{M}$ is Hamiltonian.

Now that we have the phase space, our next task is to choose a Hamiltonian.
Matrix models whose Hamiltonians are given by traces of polynomials 
in $X$ and $P$ are the most common ones in physics and geometry. Such
trace Hamiltonians are invariant under the action of $U(N)$ and
consequently the moment maps $\mu_{a}$ are conserved quantities. In what
follows we will assume that the dynamics is $U(N)$ invariant; thus the Hamiltonian will be taken as a trace Hamiltonian.

Finally let us note that the Liouville measure on $\mathcal{M}$ is
given by
\begin{equation}
    \frac{\omega^{n}}{n!}=d\nu(X)d\nu(P)
\end{equation}
where
\begin{equation}
d\nu(X)=\prod_{i}dX_{ii}\prod_{r=1,2}\prod_{i<j}dX_{rij}
\end{equation} 
and similarly for $d\nu(P)$.

\section{Statistical Mechanics of Matrix Models}

For a system in equilibrium the distribution function $\rho$ of
the statistical ensemble should be a conserved quantity. Therefore
one expects $\rho$ to be a function of the conserved quantities of
the system. So in the case of matrix models we have
\begin{equation}
    \rho=\rho(\mu_{a},H).
\end{equation}
The explicit form of the equilibrium distribution can be derived
by maximizing the entropy
\begin{equation}
    S=-\int\frac{\omega^{n}}{n!}\;\rho\; \ln\; \rho
\end{equation}
while keeping the mean values of $1$ (normalization), $\mu_{a}$
and $H$ fixed. Thus effectively one maximizes the functional
\begin{equation}
    S-\int\; \frac{\omega^{n}}{n!}\;\rho\;\theta-\int\;
    \frac{\omega^{n}}{n!}\;\rho\;
    \phi^{a}\mu_{a}-\int\frac{\omega^{n}}{n!}\;\rho\;\beta H
\end{equation}
where $\theta$, $\phi^{a}$ and $\beta$ are the Lagrange
multipliers. The result unit normalization is
\begin{equation}
    \rho=\frac{1}{Z}e^{-\phi^{a}\mu_{a}-\beta H}
\end{equation}
where $Z$ is the partition function
\begin{equation}
    Z=\int\frac{\omega^{n}}{n!}\;e^{-\phi^{a}\mu_{a}-\beta H}.
\end{equation}
The same result can also be derived by counting microstates. This
alternative derivation is given by Adler and Horwitz in
\cite{Adler-Horwitz}.

Using the explicit form of the moment maps and defining the
hermitean matrix $\phi=\phi^{a}T_{a}$ we have
\begin{equation}
    Z=\int e^{\omega-Tr\phi[X,P]-\beta H}.
\end{equation}
As we will see in the next section the combination
\begin{equation}
    \overline{\omega}=\omega-\phi^{a}\mu_{a}=\omega-Tr\phi[X,P]
\end{equation}
appearing in the partition function has an important geometric
meaning and is called the equivariant symplectic form.

From a field theoretic point of view the Lagrange multiplier
$\phi$ plays the role of an external source coupled to a conserved
charge of the system. In our model we will assume that this
external source is itself random with a Gaussian distribution
\footnote{A more familiar example of a situation where the source
term is random would be a magnetic system coupled to a random
external magnetic field.}. Thus the modified partition function of the
system with random external source is
\begin{equation}
    Z=\int \prod_{a}d\phi^{a} \int e^{\overline{\omega}-\beta H-\epsilon Tr\phi^{2}}
\end{equation}
where 
\begin{equation}
e^{-\epsilon
Tr\phi^{2}}=e^{-\frac{\epsilon}{2}\sum_{a}\phi^{a}\phi^{a}}
\end{equation} 
is the Gaussian distribution of the source $\phi$.  Note that
$Tr\phi^{2}=\frac{1}{2}\sum_{a}\phi^{a}\phi^{a}$ is the
Cartan-Killing form on $u(N)$.

\section{Equivariant Cohomology and Nonabelain localization}

In this section we review the basics of Witten's nonabelain
localization principle. Let
$(\mathcal{M},\omega)$ be a symplectic manifold and $G$ a compact
Lie group with a semi-simple Lie algebra $\mathcal{G}$. Suppose
$G$ acts on $\mathcal{M}$ with a Hamiltonian action i.e. assume
that the action is symplectic, the fundamental vector fields are
Hamiltonian and the moment maps form a Lie algebra (with respect
to the Poisson bracket) homomorphic to $\mathcal{G}$. The
fundamental vector field corresponding to an element $\xi \in
\mathcal{G}$ will be denoted by $V(\xi)$. If $\{e_{a}\}$ is a
basis for $\mathcal{G}$ then the dual basis will be denoted by
$\{e^{a}\}$ and as a short hand notation $V_{a}=V(e_{a})$.
Similarly the Lie derivative along the direction of $V_{a}$ and
the contraction with $V_{a}$ will be denoted by $\pounds_{a}$ and
$i_{a}$ respectively.

Consider the tensor product $\Omega(\mathcal{M})\otimes
P(\mathcal{G}^{*})$ where $\Omega(\mathcal{M})$ is the exterior
algebra on $\mathcal{M}$ and $P(\mathcal{G}^{*})$ is the
polynomial algebra on $\mathcal{G}$. A generic element in
$\Omega(\mathcal{M})\otimes P(\mathcal{G}^{*})$ is of the form
\begin{equation}
    \sum_{k}\; \alpha^{k}P_{k}(\phi)
\end{equation}
where $\alpha^{k}$ is a k-form on $\mathcal{M}$ and $P_{k}(\phi)$
is a polynomial in $\mathcal{G}$. The elements of
$\Omega(\mathcal{M})\otimes P(\mathcal{G}^{*})$ can be multiplied
as
\begin{equation}
\sum_{k}\; \alpha^{k}P_{k}(\phi)\sum_{l}\;
\beta^{l}Q_{l}(\phi)=\sum_{k}\sum_{l}(\alpha^{k}\wedge
\beta^{l})P_{k}(\phi)Q_{l}(\phi)
\end{equation}

Since any polynomial is a linear combination of monomials a
generic element in $\Omega(\mathcal{M})\otimes P(\mathcal{G}^{*})$
can be written as a linear combination of the elements of the form
\begin{equation}
\alpha^{p}\phi^{a_{1}}\ldots \phi^{a_{r}}.
\end{equation}
This also defines a grading in $\Omega(\mathcal{M})\otimes
P(\mathcal{G}^{*})$
\begin{eqnarray}
  \Omega(\mathcal{M})\otimes P(\mathcal{G}^{*})&=&\bigoplus_{k}A_{k}  \\
 A_{k}A_{l}&\subset& A_{k+l}\\
 A_{k}&=&span\;\{\alpha^{p}\phi^{a_{1}}\ldots \phi^{a_{r}}:p+2r=k\}
\end{eqnarray}
With this grading the multiplication is supercommutative.

There is an action of $\mathcal{G}$ on $\Omega(\mathcal{M})\otimes
P(\mathcal{G}^{*})$ given by
\begin{equation}
    \pounds_{\xi}\otimes 1+1\otimes C_{\xi}
\end{equation}
where $\pounds_{\xi}$ is the Lie derivative along the fundamental
vector field $V(\xi)$ and $C_{\xi}$ is the coadjoint action of
$\mathcal{G}$ on $P(\mathcal{G}^{*})$ given by
\begin{equation}
    C_{\xi}(\phi^{a_{1}}\ldots \phi^{a_{r}})=\sum_{n=1}^{r}\phi^{a_{1}}\ldots
    \phi^{a_{n-1}}(-\xi^{a}c_{ab}^{a_{n}}\phi^{b})\phi^{a_{n+1}\ldots\phi^{a_{r}}}.
\end{equation}
Here $c_{ab}^{c}$ are the structure constants of $\mathcal{G}$. 
Also note that $C_{\phi}(\phi^{a_{1}}\ldots \phi^{a_{r}})=0$.

The Cartan differential on $\Omega(\mathcal{M})\otimes
P(\mathcal{G}^{*})$ is defined as
\begin{equation}
    D=d\otimes 1-i_{a}\otimes \phi^{a}.
\end{equation}
In accordance with the isomorphism $P(\mathcal{G}^{*})\cong S(\mathcal{G})$ we assume that under a change of basis
of $\mathcal{G}$ the variables $\phi^{a}$ transform as the elements of the dual basis. Then it is easy to see that the combination $i_{a}\otimes \phi^{a}$ is basis
independent. 

It is also straightforward to show that $D$ is a superderivation of degree $1$
\begin{eqnarray}
   DA_{k}&\subset&A_{k+1}  \\
   D(ab)&=&(Da)b+(-1)^{k}a(Db)
\end{eqnarray}
for $a \in A_{k}$.

\newtheorem{sub2}{Proposition}[section]

\begin{sub2}
   \begin{equation}
    D^{2}=-\pounds_{a}\otimes \phi^{a}
\end{equation}
\end{sub2}
\textit{Proof:}
\begin{eqnarray}
D^{2}&=&(d\otimes 1-i_{a}\otimes \phi^{a})(d\otimes 1-i_{b}\otimes \phi^{b})\\
&=& d^{2}\otimes 1-di_{a}\otimes\phi^{a}-i_{a}d\otimes\phi^{a}+i_{a}i_{b}\otimes \phi^{a}\phi^{b}\\
&=& -\pounds_{a}\otimes\phi^{a}
\end{eqnarray}

The $G$-invariant part
$(\Omega(\mathcal{M})\otimes P(\mathcal{G}^{*}))^{G}$ of
$\Omega(\mathcal{M})\otimes P(\mathcal{G}^{*})$ is called the Cartan complex and is denoted by $\Omega_{G}(\mathcal{M})$. In other words $\Omega_{G}(\mathcal{M})$ is the set of all elements of $\Omega(\mathcal{M})\otimes P(\mathcal{G}^{*})$ which
get annihilated by $\pounds_{\xi}\otimes 1+1\otimes C_{\xi}$ for all $\xi \in \mathcal{G}$.

\begin{sub2}
\begin{equation}
(\pounds_{a}\otimes 1+1\otimes C_{a})D=D(\pounds_{a}\otimes 1+1\otimes C_{a})
\end{equation} 

\end{sub2}
\textit{Proof:}
\begin{eqnarray}
\left[ \pounds_{a}\otimes 1+1\otimes C_{a},D\right] &=&-\left[\pounds_{a},i_{b} \right]\otimes \phi^{b}-i_{b}\otimes (C_{a}\phi^{b}-\phi^{b}C_{a}) \\
&=& -c_{ab}^{c}(i_{c}\otimes\phi^{b}-i_{b}\otimes(-c_{ac}^{b}\phi^{c}) \\
&=&  -c_{ab}^{c}(i_{c}\otimes\phi^{b}+i_{c}\otimes(c_{ab}^{c}\phi^{b})=0
\end{eqnarray}

As an immediate consequence of this result we see that the Cartan
complex inherits the grading in $\Omega(\mathcal{M})\otimes
P(\mathcal{G}^{*})$
\begin{equation}
    (\Omega(\mathcal{M})\otimes
    P(\mathcal{G}^{*}))^{G}=\bigoplus_{k}A_{k}^{G}
\end{equation}
where $A_{k}^{G}$ is the $G$-invariant part of $A_{k}$. 
\begin{sub2} On $\Omega_{G}(\mathcal{M})$
\begin{equation}
D^{2}=0. 
\end{equation} 

\end{sub2}
\textit{Proof:}
 Let $b \in \Omega_{G}(\mathcal{M}) $ then
\begin{eqnarray}
 (\pounds_{a}\otimes 1+1\otimes C_{a})b=0&\Rightarrow & (1\otimes \phi^{a})(\pounds_{a}\otimes 1+1\otimes C_{a})b=0 \Rightarrow \\
 &\Rightarrow & (\pounds_{a}\otimes \phi^{a}+ 1\otimes \phi^{a}C_{a})b=0 \Rightarrow\\
 &\Rightarrow & (-D^{2}+1\otimes \phi^{a}C_{a})b=0 
\end{eqnarray} 
but as we noted earlier $(1\otimes \phi^{a}C_{a})P(\phi)=0$. So $G$-invariance of $b$ implies $D^{2}=0$ on $\Omega_{G}(\mathcal{M})$.

This result allows us to construct the cohomology of $D$ on the Cartan complex.
The resulting cohomology is called the equivariant cohomology of $\mathcal{M}$ with respect to the group $G$ and is denoted by $H_{G}(\mathcal{M})$.

Now we have to show that all this construction is not vacuous by giving an example of
an equivariant form. The form of degree 2 defined by
\begin{equation}
\overline{\omega}=\omega-\phi^{a}\mu_{a}.
\end{equation} 
is called the equivariant symplectic form. Here one can easily show that the term  $\phi^{a}\mu_{a}$ is basis independent. It is again easy to show \cite{GS} that 
$\overline{\omega}$ is $G$ invariant and $D\overline{\omega}=0$.
This equivariant symplectic form is encountered quite frequently in applications of equivariant localization principle to problems in physics and geometry (see for instance \cite{Witten1, Atiyah, Akant}).

Equivariant forms can be integrated as
\begin{equation}
    \int \; \sum_{k}\; \alpha^{k}P_{k}(\phi)=\int_{\mathcal{G}} d\phi^{1}\ldots
    d\phi^{N}e^{-\epsilon(\phi,\phi)}P_{k}(\phi) \int_{\mathcal{M}}\alpha^{k}
\end{equation}
where $(\;,\;)$ is the Cartan-Killing form on $\mathcal{G}$ and is
assumed to be nondegenerate i.e. $\mathcal{G}$ is assumed to be
semi-simple.  The exponential factor $e^{-\epsilon(\phi,\phi)}$ is
introduced to the definition as a convergence factor for the
integral over the Lie algebra. Note that $\epsilon(\phi,\phi)$ is
an equivariant form of degree 4. The integral of $\alpha^{k}$ over
$\mathcal{M}$ vanishes unless $k=dim \; \mathcal{M}$.

\begin{sub2}
If
$\mathcal{M}$ is without boundary then the integral
of any $D$-exact form vanishes.
\end{sub2}
\textit{Proof:}
Note that
\begin{equation}
    D\;\sum_{k}\; \alpha^{k}P_{k}(\phi)=\sum_{k}\;
    (d\alpha^{k})P_{k}(\phi)-(i_{a}\alpha^{k})\phi^{a}P_{k}(\phi).
\end{equation}
Then the integral of the first term over $\mathcal{M}$ vanishes
because of Stokes' theorem, while the integral of the second term
vanishes since exterior degree of $i_{a}\alpha^{k}$ is always less
than $dim \;\mathcal{M}$.

\begin{sub2}
Let $a$ be a closed equivariant form and $\lambda$ be an
equivariant 1-form. Then
\begin{equation}
    \int a=\int a\;e^{itD\lambda}
\end{equation} 

\end{sub2}
\textit{Proof:}
The proof is based on the observation that the difference of the two integrands is $D$-exact
\begin{eqnarray}
   a(1-e^{itD\lambda})&=&(1-e^{itD\lambda})a\\
   &=&-(itD\lambda+\frac{1}{2}(it)^{2}D\lambda
   D\lambda+\ldots)a\\
   &=&D\left[-(it\lambda+\frac{1}{2}(it)^{2}+\lambda
   D\lambda+\ldots)a\right]\\
   & &+(it\lambda+\frac{1}{2}(it)^{2}+\lambda
   D\lambda+\ldots)Da\\
   &=&D\left[-(it\lambda+\frac{1}{2}(it)^{2}\lambda
   D\lambda+\ldots)a\right].
\end{eqnarray}
Now the desired result follows as a simple corollary to the previous proposition.
 
 The localization principle follows from the evaluation of the
integral
\begin{equation}
\int a\;e^{itD\lambda}=\int
a\;e^{itd\lambda}e^{-it\phi^{a}\lambda(V_{a})}
\end{equation}
in the asymptotic limit $t\rightarrow\infty$. In this limit the
integral can be calculated by the method of stationary phase
\cite{GS2} which implies the localization of the integral
on the critical points of the function
\begin{equation}
\phi^{a}\lambda(V_{a}).
\end{equation}
There are two sets of equations for the critical points:
\begin{eqnarray}
\lambda(V_{a})&=&0 \\
\phi^{a}d\lambda(V_{a})&=&0.
\end{eqnarray} 
In order
to ensure the compactness of the critical point set we will assume that the last equation has the unique solution $\phi^{a}=0$.

In general there is no restriction in the choice of the
equivariant 1-form $\lambda$. We can use this freedom to our
advantage and choose a convenient $\lambda$ whose critical points
are (relatively) easy to find. In most physical and geometric
applications such a convenient choice is given by
\begin{equation}
    \lambda=JdI.
\end{equation}
Here $J$ is a $G$-invariant almost complex structure compatible
with the symplectic form $\omega$: $\pounds_{a}J=0$,
$\omega(JX,JY)=\omega(X,Y)$ and $g(X,Y):=\omega(X,JY)$ is a
positive definite metric on $M$; and
\begin{equation}
    I=\sum_{a}\mu_{a}^{2}.
\end{equation}

\begin{sub2}
$\lambda$ is $G$ invariant.

\end{sub2}
\textit{Proof:}
\begin{equation}
\pounds_{a}JdI = (\pounds_{a}J)dI+2J\pounds_{a}(\mu_{b}d\mu_{b}).
\end{equation}
The first term on the right hand side vanishes since $J$ is assumed to be a $G$-invariant. For the second term on the right hand side we have
\begin{eqnarray}
2J\pounds_{a}(\mu_{b}d\mu_{b})&=&2J((\pounds_{a}\mu_{b})d\mu_{b}+\mu_{b}d(\pounds_{a}\mu_{b})\\
                            &=&2J(\left\lbrace  \mu_{a},\mu_{b}\right\rbrace d\mu_{b}+ \mu_{b}d\left\lbrace  \mu_{a},\mu_{b}\right\rbrace\\
                &=& 2Jd(\left\lbrace  \mu_{a},\mu_{b}\right\rbrace d\mu_{b})\\
                &=& 2Jd(if_{abc}\mu_{c}\mu_{b})=0
\end{eqnarray}
since the structure constants $if_{abc}$ of a compact,
semi-simple group are completely antisymmetric.

The following result is very important in the practical applications of nonabelian localization. Its proof is
based on the compatibility of $J$ with $\omega$ and is given in \cite{Witten1}.
\begin{sub2}

The critical points of the function $ \phi^{a}\lambda(V_{a})$
coincide with the critical points of the function $I$.
\end{sub2}

Thus there are two types of critical points:
\begin{itemize}
    \item Ordinary critical points p: $\mu_{a}(p)=0$ for all $a$
    \item Higher critical points p: $dI_{p}=0$ but not all $\mu_{a}(p)$'s
    vanish.
\end{itemize}

The localization principle can now be applied to the integral of $e^{\overline{\omega}}$. 
\begin{equation}
\int e^{\overline{\omega}}=\int e^{\overline{\omega}+itD\lambda}
\end{equation}  
and implies that the integral is localized on the critical point set of $I$.

\section{Localization for Matrix Models}

Now consider the partition function of the Hermitean matrix model in the high temparature limit 
$\beta\rightarrow 0$
\begin{equation}
    Z=\int \prod_{a}d\phi^{a} \int_{\mathcal{M}} e^{\overline{\omega}-\epsilon Tr\phi^{2}}
\end{equation}
But as we noted earlier the term
\begin{equation}
Tr\phi^{2}
\end{equation} 
is nothing but the Cartan-Killing form of $u(N)$. This integral is clearly of the form discussed in the previous section.
Therefore we conclude that the integral is localized on the critical point set of 
\begin{eqnarray}
I=\sum_{a}\mu_{a}^{2}&=&\sum_{a} Tr \; (T_{a}[X,P])\; Tr \; (T_{a}[X,P])\\
&=&\frac{1}{2}\;Tr\;  [X,P]^{2}.
\end{eqnarray} 
The critical points are given by the equation
\begin{equation}
dI=Tr\; [\;P,[X,P]\;]dX-Tr\; [\;X,[X,P]\;]dP=0
\end{equation} 
which implies
\begin{eqnarray}
\left[\; P,\left[ X,P\right]\; \right] &=&0 \\
\left[\; X,\left[ X,P\right] \;\right]&=&0.
\end{eqnarray} 
Let $A_{1}\equiv X$, $A_{2}\equiv P$ and $\left[A_{\mu}\otimes A_{\nu} \right]=A_{\mu}\otimes A_{\nu}-A_{\nu}\otimes A_{\mu} $.

Now the tensor algebra generated by $span\left\lbrace A_{1},A_{2} \right\rbrace$ modulo the two sided ideal generated by
\begin{equation}
\left[ A_{\mu}\otimes \left[ A_{\mu}\otimes A_{\nu}\right]\right] 
\end{equation}  
is an example of a Yang-Mills algebra in 2 (euclidean) dimensions. This 2 dimensional Yang-Mills algebra is also an Artin-Schecter algebra and is known to be isomorphic to the universal enveloping algebra of the Heisenberg algebra \cite{Nekrasov, Connes-Violette, Artin, Artin2}.
In particular this means that the only solution for $dI=0$ is the canonical commutation relation
\begin{equation}
[X,P]=i\alpha \textbf{1}
\end{equation} 
where $\alpha$ is a real number and $\textbf{1}$ is the unit matrix. 

Thus we conclude that at high temperatures the coordinates of the phase space of a Hermitean matrix model obey the canonical commutation relations. 

The ordinary critical points correspond to the commutative case $\alpha=0$. In this case localization is onto the Cartan subalgebra of $u(N)$. In fact at finite $N$ the ordinary critical points are the only critical points since
there are no finite dimensional representations of the Heisenberg algebra. The higher critical points ($\alpha\neq 0$) exist only in the large $N$ limit. 

\paragraph{Acknowledgement:} The author would like to thank C. Deliduman, O. T. Turgut and K. Ulker for helpful conversations.

\end{document}